\documentstyle[aps,multicol,epsfig]{revtex}

\newcommand{\vn}{\vec{n}}
\newcommand{\vm}{\vec{m}}
\renewcommand{\vr}{\vec{r}}
\newcommand{\vR}{\vec{R}}
\newcommand{\vS}{\vec{S}}
\begin{document}
\title{Magnetic Vortices in High Temperature Superconductors}
\author{Per Hedeg\aa rd}
\address{\O rsted Laboratory, Niels Bohr Institute for APG, \\
Universitetsparken 5, DK-2100 Copenhagen \O\ Denmark}
\date{February 2., 2000}
\maketitle
\begin{abstract}
It is suggested that modes, observed in recent neutron scattering
experiments by Lake {\it et al.}, on La$_{2-x}$Sr$_x$CuO$_4$ in
strong magnetic fields ($\approx$ 7 T), are due to the existence
of antiferromagnetic moments associated with the cores of vortices
generated by the field. These moments form one-dimensional chains
along the $c$-axis (the vortex axis), which at finite temperatures
are disordered. At temperatures higher than 10 K the correlation
length gets shorter than the lattice parameter, resulting in no
scattering from coherent spin-waves above that temperature. The
bandwidth of the spin-waves is estimated to be $\approx$ 4 meV in
accordance with the observations.
\end{abstract}
\pacs{PACS numbers: 74.60.-w, 74.72.-h, 75.25.+z}
\begin{multicols}{2}
The superconducting and the antiferromagnetic phases are close in
the copper oxide supersuperconductors. They are close in the
overall phasediagram, where at low temperatures the
antiferromagnetic (AF) order is replaced by a superconducting (SC)
order upon doping and in certain inhomogenous phases, such as the
striped phases, the two order parameters will be literally close
geometrically. In the original observations of stripes in cuprates
by Tranquada {\it et al.}\cite{tranquada} it was found that the
stripes are in fact insulating regions with antiferromagnetic
order in a sea of superconducting material. An alternative way of
suppressing the superconductivity is to apply an external magnetic
field which will generate vortices, who's cores are
non-superconducting. It has been suggested by several, that there
is a possibility, that the cores are in fact small islands of
insulating antiferromagnets.

So far we have only indirect evidence, that such a possibility is
realized. E.g.~Scanning Tunneling Microscope (STM)
measurements\cite{renner} indicates an insulating behavior of
vortex cores in the Bismuth based cuprates. Recent inelastic
neutron scattering experiments\cite{lake} on optimally doped
La$_{2-x}$Sr$_x$CuO$_4$ in an applied magnetic field has revealed
modes with energies around 3.5 meV below 10 K. In this paper it
will be argued that such results are to be expected if in fact the
vortex cores support antiferromagnetically ordered spins.

In the experiments by Lake {\it et al.}\cite{lake} on optimally
doped La$_{2-x}$Sr$_x$CuO$_4$ below $T_c$ it is found that without
a field the spin excitations have a gap $\Delta$ = 6.7 meV, for
all measured momentum transfers. A simple BCS model for $d$-wave
superconductivity\cite{scalapino}, leads one to expect that the
gap in the spin-excitations should vanish for momentum transfers
close to the difference between two nodes in the one particle
spectrum. This is not found experimentally, from which one
concludes, that in strongly correlated systems the spin
excitations are not necessarily associated with the charge
excitations. We shall have nothing further to say about this
general problem. When a field is applied, however, the situation
changes. Then new excitations are found below a temperature $T_0$
= 10 K. These excitations have energies in the gap associated with
the field free superconductor. In fact the results can be
summarized by saying that there is a single mode with energy 3.5
meV and width 3.8 meV, although this is not the physical picture
emerging from the considerations in this paper. It is natural to
associate the excitations with the vortices generated by the
magnetic field, and Lake {\it et al.} also establishes that the
intensity of the scattering scales with the strength of the
external field, consistent with such an assignment.

Using the generalized Ginzburg-Landau model, which includes both
the AF order parameter and the SC orderparameter, and which has an
approximate SO(5) symmetry, Arovas {\it et al.}\cite{arovas} has
suggested the that vortex cores can be insulting antiferromagnets.
The scenario is not only possible in an SO(5) model, and more
recently it has also been proposed in models with spin charge
separation.\cite{lee} In this paper we shall make a minimal
assumption, that in the vortex core the AF order parameter is
non-zero in the core, and described by an envelope function
$\phi(r)$, who's total weight $\int \phi(r)^2 r dr/a^2 = N$, where
$N$ then can be interpreted as the total number of spins in the
vortex core. The excitation of such an AF vortex core has been
studied in the work by Hallundb\ae k {\it et
al.}\cite{hallundbaek}. Neglecting for a moment the interaction
between spins in different copper-oxide layers and spin
anisotropies, the spins in the core has a zero-energy, Goldstone
mode, which has excitation energy zero, corresponding to a rigid
rotation of all the spins in the core. This mode is model
independent. Since the group of spins has a finite size, the
magnons will have a discrete spectrum, and the gap to the lowest
excitation is approximately $J a/\xi$, where $J$ is the exchange
interaction in the plane ($\approx 130 meV$), $a$ is the lattice
parameter, and $\xi$ is the coherence length, i.e.~the size of the
vortex core, hence the energy is much larger than the few meV we
are considering in this paper, and we can take the spins to form a
rigid group described by one single vector $\vn$. Different models
have different other excitations of the core. E.g.~will the SO(5)
model have a resonance close to 40 meV, which describes a rotation
of the AF order parameter into the SC plane. The details are not
important
--- here we only need to know that the energy is so high, that
such modes can be neglected in the present context. There are two
final interactions, which have low energies: the Zemann coupling
to the magnetic field associated with the vortex, and the
interplane exchange coupling $J'$. Knowing the London length one
can easily estimate the Zemann energy, and this is 1 $\mu$eV,
which is way to small to be of relevance\cite{hallundbaek}. This
leaves us with $J'$. It is known to be $J' = 3.7\cdot 10 ^{-5} J$,
which also seems small, but it is an interesting feature of spin
waves in a system with different exchange constants in different
spatial directions, that modes moving along the direction with
$J'$ (the $c$-direction in our case) have energies $\omega_k =
\sqrt{JJ'(1-\cos k c)}$, i.e.~a typical energy $\sqrt{JJ'}\approx
2 $meV, which is the energy scale of the experiment.

So, we are suggesting, that what is seen in the experiment is spin
waves traveling up and down the vortex core column. From general
principles we know that a 1-dimensional antiferromagnet does not
order at any finite temperature, so we have to analyze a thermally
disordered antiferromagnet. This is most easily done using the
field theory representation for an antiferromagnet, namely the
non-linear sigma model. We start with the model in the
2-dimensional copper-oxide planes. The (imaginary time) action is
given by
\begin{eqnarray}
S_{plane} &=& \sum_i\int_0^{\beta}d\tau\int d^2 x\nonumber\\
&&\frac{1}{2}\left(\frac{1}{4aJ} \left(\frac{\partial
\vm_i}{\partial\tau}\right)^2 + a J s^2 (\nabla\vm_i)^2 +
L_{sc}\right ),
\end{eqnarray}
where $s$ is the size of the spins, $a$ is the in-plane lattice
constant and $i$ is a plane index. The last term describes the
coupling to the SC orderparameter. The precise nature of that is
not specified, except, that we assume, that in the core of the
vortex, the AF order parameter $\vm_i(\vr)$ is a constant vector
$\vn_i$, i.e.~we make the ansatz $\vm_i(\vr)= \vn_i \phi(\vr)$,
where $\phi(\vr)$ is 1 in the center of the core, and decays to
zero outside the core. The actual expectation value of the
spin-operator is then $\langle \vS_i(R_j) \rangle = s \vn_i
\phi(\vR_j) (-1)^j$, where the factor $(-1)^j$ denotes the AF
staggering, +1 on one sublattice and -1 on the other sublattice.

Let us now introduce the inter plane coupling:
\begin{equation}
L_{ip} = J' \sum_i \int d^2r \vS_i(\vr)\cdot \vS_{i+1}(\vr),
\end{equation}
which can be rewritten in terms of the in-plane AF order parameter
(and making the continuum approximation):
\begin{equation}
L_{ip} = \frac{J's^2 c N}{4} \int dz \left ( \frac{\partial
\vn}{\partial z}\right )^2,
\end{equation}
resulting finally in the effective action
\begin{equation}
S_{eff} = \frac{1}{2}\int_0^\beta d\tau \int dz \left( \chi \left
( \frac{\partial \vn}{\partial \tau}\right )^2 + \rho\left(
\frac{\partial \vn}{\partial z}\right )^2 \right ),
\end{equation}
with $\chi = N/(4 a J)$ and $\rho = J' s^2 c N/4$.

Both the ``mass'' and the ``spring constant'' scales with the size
of the vortex core, $N$, but at zero temperature this model have
low energy spin wave modes with a dispersion $i\omega_{n} =
s\sqrt{JJ'a/c}\;kc$, independent of $N$ --- and equal to the
dispersion of spin waves in an 3-D model with anisotropic exchange
constant. The size of the vortex core can be estimated from recent
measurements of $H_{c2}$ by Ando {\it et al.}\cite{ando} They find
$H_{c2}\approx 62$T, which result in an estimate for $N\approx
216$.

At finite temperatures the system is disordered. The basic
physical picture put forward in this paper, is that only
propagating modes with a wavelength shorter than the correlation
length will show up as peaks in inelastic neutron scattering.
Below a certain temperature (10 K in the experiment) the
correlation length becomes sufficiently long, that modes with wave
vectors in the experimental window ($\approx$ 50 \% of the
$c$-axis Brillouin zone) contribute to the scattering. Such modes
has energies in the upper half of the spin wave spectrum.

Accordingly we need to find the correlation length of the spins.
We use the renormalization group equations discussed extensively
by Chakravarty  {\it et al.}\cite{chakravarty}. They are for the
1-D case
\begin{eqnarray}
\frac{d g}{dl} &=& \frac{g^2}{2\pi} \coth\left(\frac{g}{2t}\right
) \\ \frac{dt}{dl} &=& t +
\frac{gt}{2\pi}\coth\left(\frac{g}{2t}\right ),
\end{eqnarray}
where $g$ is a dimensionless coupling constant:
\begin{equation}
g = \frac{1}{\sqrt{\chi\rho}} =
\frac{4}{sN}\sqrt{\frac{J}{J'}\frac{a}{c}}
\end{equation}
and $t$ is a dimensionless temperature:
\begin{equation}
t = \frac{c k_B T}{\rho} =\frac{4 k_B T}{NJ's^2}.
\end{equation}
The correlation length for the spin system, $\xi_s$ is determined
by the requirement that the renormalized length, $\xi_s(l) = \xi_s
e^{-l}$ is equal to the lattice parameter $c$\cite{chakravarty}.
This is essentially obtained when the scaling parameter, $l$ has a
value such that the renormalized temperature is $2\pi^2$.

The renormalization equations can be solved exactly, and the
solution is
\begin{eqnarray}\label{losning}
t(l) = \frac{t_0 e^l}{1-g_0 f(l)}, \qquad\mbox{where} \nonumber\\
f(l) = \int_0^l \frac{dl'}{2\pi}
\coth\left(\frac{g_0}{2t_0}e^{-l'}\right).
\end{eqnarray}
The correlation length is finite at all temperatures, with a
limiting value
\begin{equation}
\xi_s(T=0) = c e^{2\pi/g_0}.
\end{equation}

For the parameters relevant to our system, the scaling starts at
values
\begin{eqnarray}
g_0 &\approx& 5 \\ t_0 &\approx& 1.7 \;T ,
\end{eqnarray}
where $T$ is the actual temperature measured in Kelvin. These are
approximate values derived from very difficult experiments
($H_{c2}$ and $J'$ are hard to get), so precise scale for e.~g.
$t_0$ can easily vary with a factor of 2. Chosing the temperature
scale factor to be 3.4 instead of 1.7, which corresponds to the
product $N J'$ being a factor 2 smaller than quoted above, we get
the curves shown in Figure 1 for the correlation length $\xi(T)$
as a function of temperature and for 3 values, 1, 5, 10 for $g_0$.
We see that a strong growth of correlation length around 10 Kelvin
is quite compatible with known values for the size of the vortex
core and the strength of the interplane exchange coupling.

\begin{figure}
\centerline{\epsfig{file=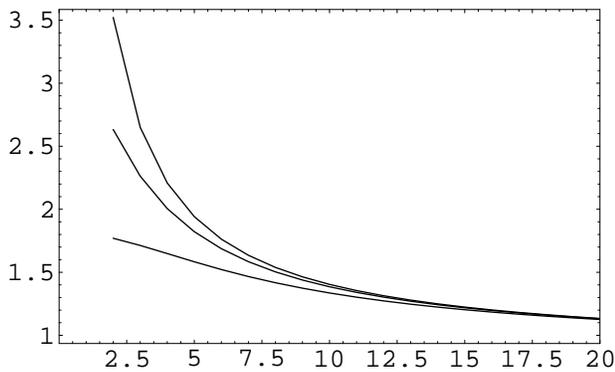}} \narrowtext
\caption{ The correlation length of spins in the vortex core along
the $c$-axis as a function of temperature. The lower curve is for
$g_0=10$, the middle for $g_0=5$ and the upper for $g_0=1$. The
temperatur scale is given by $t_0 = 3.4 T$, where $T$ is measured
in Kelvin. }
\end{figure}

In conclusion, we have proposed an explanation of the recent
experimental finding by Lake {\it et al.} of magnetic scattering
from optimally doped La$_{2-x}$Sr$_x$CuO$_4$ in a strong magnetic
field, below 10 Kelvin at an energy of approximately 3 meV. This
energy scale is the natural one for spin waves in a system of
antiferromagnetically ordered spins residing in the vortex cores
of the superconductor. This being a 1-dimensional system no long
range order exist at any finite temperature, so inelastic
scattering creating reasonably longlived spin waves requires a
correlation length larger than the interplane distance. We have
shown that for reasonable values of the vortex core size and the
interplane exchange coupling, this happens for temperatures less
than 10 Kelvin. It is beyond the scope of this paper to give a
detailed calculation of the actual theoretical neutron scattering
spectra, but calculations of this is underway.

I thank Bella Lake and Kim Lefmann for an early peek at their
data.

\end{multicols}

\begin{references}
\bibitem{lake}B.~Lake, G.~Aeppli, K.~N.~Clausen, D.~F.~McMorrow,
K.~Lefmann, N.~E.~Hussey, N.~Mangkorntong, N.~Nohara, H.~Takagi,
T.~E.~Mason, and A.~Schr\"oder, submitted to Science.
\bibitem{tranquada} J.~M.~Tranquada, B.~J.~Sternlib, J.~D.~Axe,
Y.~Nakamura and S.~Uchida, Nature, {\bf 375}, 561 (1995).
\bibitem{chakravarty}S.~Chakravarty, B.~I.~Halperin, and
D.~R.~Nelson, Phys. Rev. B {\bf 39}, 2344 (1989).
\bibitem{arovas}D.~P.~Arovas, A.~J.~Berlinsky, C.~Kallin,
S.~C.~Zhang, Phys. Rev. Lett. {\bf 79}, 2871 (1997).
\bibitem{hallundbaek}H.~Bruus, K.~A.~Eriksen, M.~Hallundb\ae k,
and P.~Hedeg\aa rd, Phys. Rev. B {\bf 59}, 4349 (1999).
\bibitem{renner}Ch.~Renner, B.~Revaz, K.~Kadowaki, I.~Maggio-Aprile, and
\O.~Fischer, Phys. Rev. Lett. {\bf 80}, 3606 (1998).
\bibitem{lee}Jung Hoon Han, Dung-Hai Lee, Phys. Rev. Lett. {\bf 85}, 1100
(2000).
\bibitem{scalapino}N.~Bulut and D.~J.~Scalapino, Phys. Rev. B {\bf 50}, 16078
(1994).
\bibitem{ando}Y.~Ando, G.~S.~Boebinger, A.~Passner, L.~F.~Schneemeyer, T.~Kimura, M.~Okuya, S.~Watauchi,
J.~Shimoyama, K.~Kishio, K.~Tamasaku, N.~Ichikawa, and S.~Uchida,
Phys. Rev. B {\bf 60}, 12475 (1999).
\end{references}
\end{document}